\documentclass[twocolumn,eqsecnum]{article}
\usepackage{epsfig}
\def\pmb#1{\setbox0=\hbox{#1}%
     \kern-.025em\copy0\kern-\wd0
      \kern.05em\copy0\kern-\wd0
       \kern-.025em\raise.0433em\box0}
\def\bomeg{\pmb{$\omega$}}

\def\blambda{\pmb{$\lambda$}}
\def\bPhi{\pmb{$\Phi$}}

\def\beq{\begin{equation}}
\def\eeq{\end{equation}}
\def\bea{\begin{eqnarray}}
\def\eea{\end{eqnarray}}
\begin{document}
%
\title{\bf Oscillations in finite Fermi systems}
\author{ V. I. Abrosimov$^{\rm a}$, A.Dellafiore $^{\rm b}$ and
F.Matera$^{\rm b}$\\
\it $^{\rm a}$ Institute for Nuclear Research, 03028 Kiev, Ukraine\\
$^{\rm b}$ \it Istituto Nazionale di Fisica Nucleare and Dipartimento
di Fisica,\\
\it Universita' di Firenze, via Sansone 1, 50019 Sesto F.no (Firenze), Italy}
\date{}
\maketitle
\begin{abstract}
A semiclassical linear response theory based on the Vlasov equation is
reviewed. The approach discussed here differs from the classical one
of Vlasov and Landau for the fact that the finite size of the system
is explicitly taken into account. The non-trivial problem of deciding
which boundary conditions are more appropriate  for the fluctuations
of the phase-space density has been circumvented by studying solutions
corresponding to different boundary conditions (fixed and moving
surface). The fixed-surface theory has been applied to systems
having both spherical and spheroidal equilibrium shapes.
The moving-surface theory is related to the
liquid-drop model of the nucleus and  from it one can
obtain a kinetic-theory description of surface and compression modes in
nuclei.
Quantum corrections to the semiclassical theory are also briefly
discussed.
\end{abstract}

\section{INTRODUCTION}

In this paper we review the work of our groups on the application of the
Vlasov kinetic equation to small-amplitude density vibrations in
{\em finite} Fermi systems.
The phenomenology that is described by this approach includes
giant resonances in nuclei, surface plasmons in atomic clusters and
possible collective excitations of the electrons in heavy atoms (which
however turned out not to be really collective). In the approach of
Jeans \cite{jea}, Vlasov \cite{vla1,vla2}
and Landau \cite{lan1,lan2} the theory has been developed
for infinite homogeneous systems,
later Kirzhnitz {\it et al.} \cite{kir} extended it to non-uniform systems.

The quantity that is studied in this approach is the phase-space density
(or single-particle distribution) $f({\bf r},{\bf p},t)$. This
quantity, when multiplied by the phase space volume $d{\bf r}d{\bf
p}$ gives the number of particles contained in that volume and has
several useful properties since its knowledge allows us to obtain
information about macroscopic properties of the system like, for
example, the ordinary density
\beq
\label{rho}
\varrho ({\bf r},t)=\int d{\bf p}f({\bf r},{\bf p},t)\,,
\eeq
or the pressure tensor
\beq
\label{pre}
\Pi_{jk} ({\bf r},t)=\int d{\bf p}p_{j}v_{k}f({\bf r},{\bf p},t)\,.
\eeq

We limit our interest to small oscillations of $f({\bf r},{\bf p},t)$
about its equilibrium value $f_{0}({\bf r},{\bf p})$, which is
supposed to describe the stationary state of the many-body system.
If the system is subject to a weak external driving potential $
V_{ext}({\bf r},t)=\beta(t) Q({\bf r})$, then $f_{0}$ is changed by the
small amount
$\delta f({\bf r},{\bf p},t)$, consequently the local equilibrium
density also changes by
\beq
\label{delrhotot}
\delta \varrho ({\bf r},t)=\int d{\bf p}\delta f({\bf r},{\bf
p},t)\,,
\eeq
and the equilibrium mean field $V_{0}({\bf r})$ by
\beq
\label{dev}
\delta V({\bf r},t)=\int d{\bf r}'v({\bf r},{\bf r}')
\delta\varrho ({\bf r}',t)\,.
\eeq
The quantity $v({\bf r},{\bf r}')$ is the effective interaction
between two constituents of the many-body system. We
shall consider only quantities at first order in $\delta f$, moreover the
equilibrium distribution will be assumed to depend only on the energy
of the particles: $f_{0}({\bf r},{\bf p})=F(h_{0}({\bf r},{\bf p}))$,
where
\beq
h_{0}({\bf r},{\bf p})=\frac{p^{2}}{2m}+V_{0}({\bf r})
\eeq
is the equilibrium hamiltonian.

 In a fully self-consistent approach
the equilibrium mean field should be given by
\beq
\label{mfi}
V_{0}({\bf r})=\int d{\bf r}'v({\bf r},{\bf r}')
\int d{\bf p}f_{0}({\bf r}',{\bf p})\,,
\eeq
however we shall use phenomenological approximations to it. There are
two kinds of self-consistency requirements: the static self-consistency
expressed by Eq. (\ref{mfi}), and the analogous dynamic
self-consistency condition (\ref{dev}). We shall treat the static
self-consistency condition more loosely, but will respect the dynamic
condition (\ref{dev}) because it is essential in the
description of collective modes.

\section{HISTORICAL \\
 REMARKS}

The following differential equation for $f({\bf r},{\bf p},t)$ was
considered by A.A. Vlasov in a study of plasma oscillations
\cite{vla1,vla2}:
\beq
\label{kin}
\frac{\partial f}{\partial t}+\frac{{\bf p}}{m}\frac{\partial
f}{\partial {\bf r}}-(\frac{\partial V}{\partial {\bf r}})\cdot
{{\partial f}\over {\partial {\bf p}}}=0\,,
\eeq
together with the Poisson equation\footnote{Vlasov actually
considered coupling to the full elecromagnetic field by using the
Maxwell equations, here we simplify his argument and assume that the
time dependence is sufficiently slow so that the laws of
electrostatics can be applied at any instant $t$}
\beq
\label{poi}
\nabla^{2} V=-4\pi e^{2}\varrho\,.
\eeq
Because of Eq.(\ref{rho}), equations (\ref{kin}) and (\ref{poi}) form a
set of coupled equations that can be solved self-consistently. Thus
the kinetic equation (\ref{kin}), supplemented by (\ref{poi}), can be
considered as a dynamical generalization of the static Thomas-Fermi
method.

Althogh Eq.(\ref{kin}) is a simplified version of an equation already
derived by Boltzmann and the idea of self-consistency in connection
with it had already been used by  Jeans (see \cite{hen} for a
discussion of hystorical priorities), we follow the use common both
in plasma and in nuclear physics, and refer to Eq.(\ref{kin}) as ''the
Vlasov equation''.

The work of Vlasov was criticized by Landau \cite{lan1} who pointed
out some mathematical
inconsistencies in Vlasov's solution and derived a rigorous solution
of Eqs. (\ref {kin}-\ref{poi}) for plasma oscillations.

In his later work on Fermi liquids \cite{lan2} Landau obtained a
kinetic equation similar to Eq.(\ref{kin}) (see also Ref. \cite{bay},
p.18)
and used it to study the propagation of zero sound in liquid helium.

Vlasov and Landau were interested in macroscopic systems,
so they did not worry about surface
effects; in their approach the system is supposed to be homogeneous
and infinite, so that translation invariance considerably simplifies
calculations. Extension of their approach to microscopic systems like
atoms \cite{kir} or nuclei \cite{ber}, apart from giving up
translation invariance, has to face the non-trivial problem of
deciding which boundary conditions are most appropriate for the
fluctuations of the single-particle distribution $\delta f({\bf r},{\bf
p},t)$.

Kirzhnitz and collaborators \cite{kir} based their search for
possible collective excitations of the electron cloud in heavy atoms
on the linearized Vlasov equation. The main results of their work can
be summarized in the two following equations for the polarization
propagator:
\bea
\label{rpa}
&&\Pi({\bf r},{\bf r}',\omega)=\Pi^{0}({\bf r},{\bf
r}',\omega)\\
&&+\int d{\bf x}\int d{\bf y}\Pi^{0}({\bf r},{\bf x},\omega) v({\bf
x},{\bf y})\Pi({\bf y},{\bf r}',\omega)\nonumber\,,
\eea
and
\bea
\label{pi0}
&&\Pi^{0}({\bf r},{\bf r}',\omega)=-\frac{mp_{F}(r')}{\pi^{2}\hbar
^{3}}{\Big (}\delta ({\bf r}-{\bf r}')\qquad\\
&&+i\omega \int_{-\infty}^{0} dt
e^{-i\omega t}<\delta ({\bf r}(t)-{\bf r}')>{\Big )}\nonumber\,.
\eea

The propagator $\Pi({\bf r},{\bf r}',\omega)$ relates the (Fourier
transformed in time) external disturbance at point ${\bf r}'$ to the density
response at point ${\bf r}$:
\beq
\label {pol}
\delta \varrho ({\bf r},\omega )=\int d {\bf r}'\Pi({\bf r},{\bf
r}',\omega)  V_{ext}({\bf r}',\omega)\,,
\eeq
while $\Pi^{0}$ is the same propagator evaluated in the static mean
field, that is, by neglecting the mean-field fluctuation induced by
the external force. The most interesting part in the expression
(\ref{pi0}) is the time integral. The quantity ${\bf r}(t)$ in the
integrand is the coordinate of a particle that, at time $t=0$ is at
point ${\bf r}$ and that moves in the static mean field according to
the laws of classical physics. The $\delta$-function in the integrand
contributes to the integral whenever ${\bf r}(t)={\bf r}'$, that is
whenever a particle in ${\bf r}$ at $t=0$ has been through ${\bf r}'$
at some earlier time. The symbol $<~>$ in Eq. (\ref{pi0}) means
averaging over the directions of the particle velocity at $t=0$, which
requires that all the possible classical trajectories going from ${\bf
r}'$ to ${\bf r}$ are taken into account when evaluating $\Pi^{0}$.
In Fig. 1 we give a schematic picture of a few classical orbits
contributing to the time integral in Eq. (\ref{pi0}).
\begin{figure}
\centerline{\psfig{figure=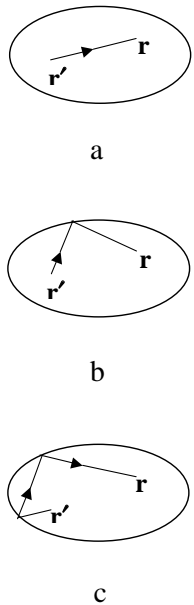,height=9.0cm}}
\caption{\small Schematic representation of a few classical orbits
contributing to
the
polarization propagator (\ref{pi0}) for a finite system. Figure (a)
shows the direct trajectory from ${\bf r}'$ to ${\bf r}$, this is the
only trajectory contributing to the propagator in a collisionless
homogeneous system. Figures (b) and (c) show two typical trajectories
with one and two reflections at the boundary.}
\end{figure}

Bertsch \cite{ber} used the Vlasov equation as a starting point for a
theory of giant resonances in nuclei. He pointed out that, since the
Vlasov dynamics preserves the density of points in phase space, it
does not violate the Pauli principle, even if it is a completely
classical theory. Thus, Eq.(\ref{kin}), after having been used by
Jeans for stellar systems, by Vlasov and Landau for plasmas, by Landau
for liquid helium and by Kirzhnitz {\it et al.} for atoms, found its
way into nuclear physics as well.

\section{FIXED BOUNDARY}
\subsection{Integrable systems}
The expression (\ref{pi0}) for $\Pi^{0}({\bf r},{\bf r}',\omega)$ is
quite general and in principle it is valid also for systems in which
the motion of particles in the equilibrium mean field is chaotic,
however it is not really suitable for practical calculations since
evaluating the time integral requires following the particle
trajectories over an infinite time interval. Kirzhnitz {\it et al.}
avoided this difficulty by restricting themselves to systems in which
the particle motion is periodic, but this is a very severe limitation.
In the more recent
work of Brink {\it et al.} \cite{bri} it has been shown that there is an
important class
of finite systems for which the solution of the linearized Vlasov equation
acquires a relatively simple form: they are all the systems for which the
motion of particles in the equilibrium mean field can be described by an
integrable Hamiltonian, this includes for example all spherically
symmetric systems, but also some deformed systems. In these systems
the particle motion is multiply periodic and, as a consequence, it is
sufficient to follow the particle motion only over a finite time
interval. The crucial step in applying the approach of \cite{kir} to
integrable systems, is an appropriate change of variables. The $({\bf
r},{\bf p})$ variables are convenient for translation-invariant systems,
since in that case ${\partial V_{0}}/{\partial {\bf r}}=0$ and the
linearized version of Eq. (\ref{kin}) becomes simpler. A similar
simplification can be obtained for integrable systems if action-angle
variables $({\bf I},{\bf \Phi})$ are used instead of $({\bf r},{\bf p})$.
For these systems the action variables $I_{\alpha}$ are constants of
the motion, while the conjugate angle variables $\Phi_{\alpha}$ are
linear functions of time (see for example Ref. \cite{gol}, p. 457).
An important property of these variables is that the motion is periodic
in the angle variables with period $2\pi$. Consequently the field felt
by a particle that is moving along such a trajectory can be Fourier expanded as
\beq
\label{fou1}
V_{ext}({\bf r},\omega)=\beta(\omega)\sum_{{\bf n}}Q_{\bf n}({\bf
I}) e^{i{\bf n}\cdot {\bf \Phi}}\,,
\eeq
where ${\bf n}$ is a three-dimensional vector with integer components.
Similar expansions hold also for the fluctuations of the mean field
\beq
\label{fou2}
\delta V({\bf r},\omega)=\sum_{{\bf n}}\delta V_{\bf n}({\bf
I},\omega) e^{i{\bf n}\cdot {\bf \Phi}}
\eeq
and of the distribution function
\beq
\label{fou3}
\delta f({\bf r},{\bf p},\omega)=\sum_{\bf n}\delta f_{\bf n}({\bf I},\omega)
e^{i{\bf n}\cdot {\bf \Phi}}\,.
\eeq

When Eq.(\ref{kin}) is Fourier transformed in time and linearized by
neglecting terms of order $(\delta f)^{2}$ and higher, it reads
\bea
\label{lki1}
&&-i\omega\delta f(\omega)+\{\delta f,h_{0}\}\\
&&+F'(h_{0})\{h_{0},[V^{ext}(\omega)+\delta V(\omega)]\}=0\nonumber\,.
\eea
Here the braces denote Poisson brackets, that can be evaluated
according to any set of canonically conjugate variables, using ${\bf
r}$ and ${\bf p}$ gives the linearized version of Eq.(\ref{kin}),
while using ${\bf I}$ and ${\bf \Phi}$ gives
\bea
\label{lki2}
&&-i\omega\delta f(\omega)+ \bomeg\cdot \frac{\partial \delta
f}{\partial{\bf \Phi}}=\\
&&F'(h_{0})\,\bomeg \cdot {\Big
(}\frac{\partial V^{ext}}{\partial {\bf \Phi}}+\frac{\partial
\delta V}{\partial {\bf \Phi}}{\Big )}\nonumber\,.
\eea
The three components of the vector $ \bomeg$ determine the
time-dependence of the angle variables:
\beq
\dot \Phi_{\alpha}=\frac{\partial h_{0}({\bf I},{\bf \Phi})}{\partial
I_{\alpha}}=\omega_{\alpha}\,,
\eeq
while, of course
\beq
\dot I_{\alpha}=-\frac{\partial h_{0}({\bf I},{\bf \Phi})}{\partial
\Phi_{\alpha}}=0\,,
\eeq
and $h_{0}=h_{0}({\bf I})$.

By using the expansions (\ref{fou1}-\ref{fou3}), Eq.(\ref{lki2}) gives
\bea
\label{sol}
&&\delta f_{\bf n}({\bf I},\omega)=F'(h_{0})
{\Big [}\beta(\omega)Q_{\bf n}({\bf I})\\
&&+\delta V_{\bf n}({\bf I},\omega){\Big ]}\frac{{\bf n}\cdot
{\bomeg}}{{\bf n}\cdot {\bomeg}-(\omega +i\varepsilon)}\nonumber\,.
\eea
We have added a vanishingly small positive imaginary part $i\varepsilon$ in
order to specify the behaviour of $\delta f_{\bf n}(\omega)$ at the
pole \footnote{In this subsection we have recalled the content of
Sect.3 of \cite{bri}. Similar results had been obtained previously in
Ref. \cite{pol}, A.D. apologises to those authors for not having been
aware of their work before}.

Equation (\ref{sol}) gives an explicit solution of the linearized Vlasov
equation only if the fluctuation of the mean field $\delta V$ can be
neglected, otherwise it gives only an implicit solution,
since $\delta V$ does depend on $\delta f$; however
its form immediately suggests an iterative procedure for obtaining
the full solution: first evaluate $\delta f^{0}$ by neglecting
$\delta V$ in (\ref{sol}), then evaluate $\delta V$ by using $\delta
f^{0}$ as input and repeat until convergence is obtained.

The quantity $\delta f$ is not very convenient for discussing the
solution of the linearized Vlasov equation, since it depends on the
external driving force, so it is appropriate to introduce the
polarization propagators (\ref{rpa}) and (\ref{pi0}) , whose properties
depend only on the system. Clearly, once we know $\delta f$, by using
Eqs. (\ref{rho}) and (\ref{pol}), we can obtain a corresponding
expression for the propagators $\Pi^{0}$ and $\Pi$. We prefer to use
their momentum representations, obtained from (\ref{rpa}) and (\ref{pi0})
by taking Fourier transforms with respect to the coordinates:
\beq
\label{qprop}
\Pi({\bf q},{\bf q}',\omega)=\int d{\bf r} \int d{\bf r}'
e^{-i{\bf q}\cdot {\bf r}}
\Pi({\bf r},{\bf r}',\omega)e^{i{\bf q}'\cdot {\bf r}'}\,.
\eeq
The expression of the $\Pi^{0}$ propagator is \cite{abm}
\bea
\label{qpi0}
&&\Pi{^0}({\bf q}^\prime,{\bf q},\omega)=\\
&&(2\pi)^3
\sum_{\bf n}\int d{\bf I}\,F'(h_{0}({\bf I}))\nonumber\\
&&\times\frac{{\bf n}\cdot {\bomeg}({\bf I})}
{{\bf n}\cdot {\bomeg}({\bf I})-(\omega +
i\varepsilon)}\,Q^{*}_{\bf n}({\bf q}',{\bf I})~Q_{\bf n}({\bf
q,I})\nonumber\,,
\eea
with the Fourier coefficients
\beq
\label{fou}
Q_{\bf n}({\bf q,I})=\frac{1}{(2\pi)^3}\int d\bPhi\,e^{-i{\bf n}\cdot{\bPhi}}\,
e^{i{\bf q}\cdot{\bf r}}\,.
\eeq

If the interaction $v({\bf x},{\bf y})$ does depend only on the
distance between particles, the integral equation (\ref{rpa}) becomes
\cite{dema}
\bea
\label{qrpa}
&&\Pi({\bf q}^{\prime},{\bf q},\omega)=\Pi^{0}({\bf q}^{\prime},{\bf
q},\omega)\\
&&+\frac{1}{(2\pi)^{3}} \int d{\bf k}\,\Pi^{0}({\bf q}^{\prime},{\bf
k},\omega)\,
v(k)\,\Pi({\bf k},{\bf q},\omega)\nonumber\,.
\eea

\subsection{Spherical systems}

Spherical systems are an important class of integrable systems, hence
they deserve a more detailed discussion.
Actually these systems are over-integrable, since in this case there are four
constants of motion (the
particle enegy $\epsilon$ and the three components of its angular
momentum $\blambda={\bf r}\times {\bf p})$, so one of the angle
variables must also be constant.
The action-angle variables $({\bf I},{\bf \Phi})$ allow for a very
compact solution of the linearized Vlasov equation (\ref{lki1}), but they are
not the only set of variables that can simplify the
problem. The important point is to choose as variables the largest
possible number of constants of motion, so that the number of partial
derivatives in Eq.(\ref{lki1}) can be reduced.
\begin{figure}
\centerline{\psfig{figure=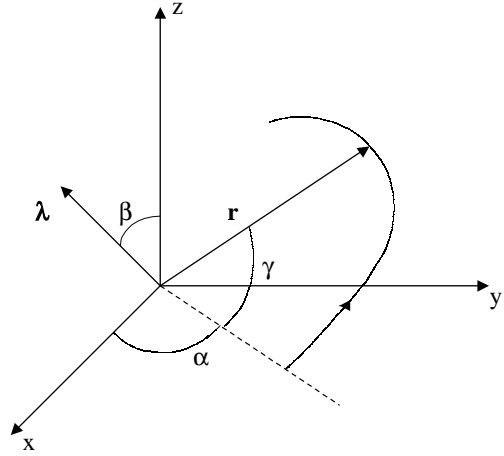,height=6.0cm}}
\caption{Angular elements of the orbit. For a particle moving in a central
force field, the angles $\alpha$ and $\beta$ are constant.}
\end{figure}
The variables used in
\cite{bri}
instead of $({\bf r},{\bf p})$ are $(\epsilon, \lambda, r, \alpha,
\beta, \gamma)$ of which four ($\epsilon$, $\lambda$, and the two angles
$\alpha$ and $\beta$ shown in Fig.2) are constants of motion. So we are
left only with the two partial derivatives with respect to $r$ and $\gamma$
in Eq.(\ref{lki1}). The $\gamma$-derivative can be eliminated by means
of the expansion
\beq
\label{pwe}
\delta f({\bf r},{\bf p},\omega)=\sum_{LM}\delta f_{LM}(r,{\bf
p},\omega) Y_{LM}(\theta,\varphi)
\eeq
and by using the well-known transformation property of the spherical
harmonics $Y_{LM}(\theta,\varphi)$ under the
rotation specified by the Euler angles $(\alpha,\beta,\gamma)$
(see e. g. \cite{bris}, p. 28):
\beq
Y_{LM}(\theta,\varphi)=\sum_{N}{\Big (}{\mathcal
D}^{L}_{MN}(\alpha,\beta,\gamma)
{\Big )}^{*} Y_{LN}(\theta',\varphi')\,.
\eeq
If the new reference frame rotates with the particle, the angles
$\theta'$ and $\varphi'$ are constant and the only time-dependent angle
is $\gamma$.
The quantities ${\mathcal D}^{L}_{MN}(\alpha,\beta,\gamma)$ are the
rotation matrices and their explicit $\gamma$-dependence, of the kind
$e^{-iN\gamma}$, can be
exploited to eliminate the $\gamma$-derivative. In this way the
initial three-dimensional problem is reduced to a one-dimensional
problem for particles moving in the effective potential
$V_{eff}=V_{0}(r)+\frac{\lambda^{2}}{2mr^{2}}$.

Then, as shown in \cite{bri}, the expansion (\ref{pwe}) can be written as
\footnote{Note that,
unlike in \cite{bri}, here we do not include the factor
$Y_{LN}(\frac{\pi}{2},\frac{\pi}{2})$ in $\delta f^{L\pm}_{MN}$.}
\bea
&&\delta f({\bf r},{\bf p},\omega)=\\
&&\sum_{LMN}{\Big[}\delta f^{L+}_{MN}(\epsilon,\lambda,r,\omega)+
\delta f^{L-}_{MN}(\epsilon,\lambda,r,\omega){\Big
]}\nonumber\\
&&\times {\Big (}{\mathcal D}^{L}_{MN}(\alpha,\beta,\gamma){\Big )}^{*}
Y_{LN}(\frac{\pi}{2},\frac{\pi}{2})\nonumber\,.
\eea
The functions $\delta f^{L\pm}_{MN}$ refer to particles having positive
$(\delta f^{L+}_{MN})$ or negative $(\delta f^{L-}_{MN})$ components of the
radial
velocity, of magnitude
$v_{r}(\epsilon,\lambda,r)=\sqrt{\frac{2}{m}(\epsilon-V_{eff})}$.

The linearized Vlasov equation (\ref{lki1}) implies the following
system of coupled first-order differential equations in the variable
$r$ for the functions $\delta f^{L\pm}_{MN}$
\beq
\label{sys}
\frac{\partial}{\partial r}\delta f^{L\pm}_{MN}\mp A_{N}\delta
f^{L\pm}_{MN}=B^{L\pm}_{MN}\,,
\eeq
with
\beq
A_{N}(\epsilon,\lambda,r,\omega)=\frac{i\omega}{v_{r}(\epsilon,\lambda,r)}
-\frac{iN}{v_{r}(\epsilon,\lambda,r)}\frac{\lambda}{mr^{2}}
\eeq
and
\bea
\label{inho}
&&B^{L\pm}_{MN}(\epsilon,\lambda,r,\omega)=\\
&&F'(\epsilon){\Big (}\frac{\partial}{\partial r}
\pm\frac{iN}{v_{r}(\epsilon,\lambda,r)}\frac{\lambda}{mr^{2}}{\Big
)}\nonumber\\
&&{\Big [}\beta(\omega)Q_{LM}(r)+\delta V_{LM}(r,\omega) {\Big
]}\nonumber\,.
\eea
The functions $Q_{LM}(r)$ and $\delta V_{LM}(r,\omega)$ are the
coefficients of a multipole expansion similar to (\ref{pwe}) for the external
driving field and for the induced mean-field fluctuations, respectively.
Because of Eq.(\ref{dev}), the term $\delta V_{LM}(r,\omega)$ couples
the two equations (\ref{sys}) for $\delta f^{L+}_{MN}$ and $\delta
f^{L-}_{MN}$:
\bea
\label{delvlm}
&&\delta V_{LM}(r,\omega)=\\
&&\frac{8\pi^{2}}{2L+1}\sum_{N=-L}^{L}
{\Big |}Y_{LN}(\frac{\pi}{2},\frac{\pi}{2}){\Big |}^{2}\nonumber\\
&&\int d\epsilon \int d\lambda \lambda \int \frac{d
r'}{v_{r}(r')}v_{L}(r,r')\nonumber\\
&&{\big [}\delta  f^{L+}_{MN}(\epsilon,\lambda,r',\omega)
+ \delta  f^{L-}_{MN}(\epsilon,\lambda,r',\omega){\big ]}\nonumber\,.
\eea

Before solving the system of equations (\ref{sys}), we must specify the
boundary conditions satisfied by the phase-space density
fluctuations. The conditions used in \cite{bri} are
\bea
\label{bc1}
&&\delta f^{L+}_{MN}(r_{1})=\delta f^{L-}_{MN}(r_{1})\,, \\
&&\delta f^{L+}_{MN}(r_{2})=\delta f^{L-}_{MN}(r_{2})\,,
\label{bc2}
\eea
where $r_{1(2)}$ are the classical turning points, for which
$v_{r}(\epsilon,\lambda,r)=0$.

With these boundary conditions the solution of the linearized Vlasov
equation for spherical systems agrees with that given by the method of
action-angle variables \cite{bri}.

There are some some further
simplifications for the polarization propagators of spherical systems,
compared to those given in the previous subsection. The
three-dimensional integral equation (\ref{qrpa}) reduces to a set of
one-dimensional integral equations for each multipolarity $L$
\bea
\label{pil}
&&\Pi_{L}( q', q,\omega)=\Pi^{0}_{L}(q',q,\omega)+\frac{1}{(2\pi)^{3}}\\
&& \int_{0}^{\infty}dk k^{2}\,\Pi^{0}_{L}(q', k,\omega)\,
v(k)\,\Pi_{L}(k,q,\omega)\nonumber\,,
\eea
and the $\Pi^{0}$ propagator also becomes somewhat simpler than
(\ref{qpi0}):
\bea
\label{spi0}
&&\Pi^{0}_{L}(q',q,\omega)=\\
&&\frac{8\pi^{2}}{2L+1}\sum_{n=-\infty}^{+\infty}
\sum_{N=-L}^{L}{\Big|}Y_{LN}(\frac{\pi}{2}\frac{\pi}{2}){\Big |}^{2}\nonumber\\
&&\int d\epsilon F'(\epsilon) \int d\lambda\,\lambda\, T
\frac{n\omega_{0}+N\omega_{\varphi}}{n\omega_{0}+N\omega_{\varphi}-
(\omega+i\varepsilon)}\nonumber\\
&&Q^{(L)\,*}_{nN}(q',\epsilon,\lambda)
Q^{(L)}_{nN}(q,\epsilon,\lambda)\nonumber\,.
\eea
Here $\omega_{0}(\epsilon,\lambda)$ is the frequency of radial motion
of a particle with enegy $\epsilon$ and magnitude of angular momentum
$\lambda$ in the effective potential $V_{eff}$,
while $\omega_{\varphi}(\epsilon,\lambda)$ is the precession
frequency of the periapsis in the plane of the orbit
(\cite{gol},p.509). These two frequencies determine the shape of the
orbit in a central potential. The period of radial motion $T$ is
$T(\epsilon,\lambda)=2\pi /\omega_{0}$. The Fourier
coefficients $Q^{(L)}_{nN}(q,\epsilon,\lambda)$ are given by
\bea
&&Q^{(L)}_{nN}(q,\epsilon,\lambda)=\\
&&\frac{2}{T}\int_{r_{1}}^{r_{2}}
\frac{dr}{v_{r}(\epsilon,\lambda,r)}\cos [s_{nN}(r)]\,
Q_{L}(qr)\nonumber\,,
\eea
with
\beq
Q_{L}(qr)=4\pi i^{L} j_{L}(qr)\,.
\eeq
The phases $s_{nN}(r)$ are
\beq
s_{nN}(r)=n\omega_{0}\tau(r)+N{\big(}\omega_{\varphi}\tau(r)-\gamma(r){\big
)}\,,
\eeq
where
\beq
\tau(r)=\int_{r_{1}}^{r}\frac{d r'}{v_{r}(\epsilon,\lambda,r')}\,,
\eeq
is the time taken by a particle to move from the inner turning point
$r_{1}$ to point $r$, and
\beq
\gamma(r)=\int_{r_{1}}^{r}\frac{d
r'}{v_{r}(\epsilon,\lambda,r')}\frac{\lambda}{mr'^{2}}
\eeq
is the angle spanned by the radius vector during the time interval
$\tau(r)$.

In Eqs. (\ref{qpi0}) and (\ref{spi0}) it has been asumed that the
equilibrium distribution of particles depends only on the particle
energy. Usually one takes $F(\epsilon)\propto
\theta(\epsilon_{F}-\epsilon)$,where $\epsilon_{F}$ is the Fermi
energy, so that $ F'(\epsilon) \propto \delta(\epsilon_{F}-\epsilon)$
and the expression (\ref{spi0}) becomes simpler.

At first sight it might seem that the infinite sum over $n$ in (\ref{spi0})
might create difficulties, but in practice it is enough to include
in the sum the very first few terms around $n=0$ to get a sufficient
approximation.

\subsection{Atomic plasmons?}

The early calculations of Kirzhnitz and collaborators
suggested the existence of two collective high-energy excitations
in the spectrum of heavy atoms  \cite{kir}. These excitations had
been interpreted as collective, plasmon-type oscillations of the
atomic electron cloud. In order to apply their formalism, Kirzhnitz
{\it et al.} had to approximate the atomic self-consistent mean field
with a potential in which electrons having zero total energy move
along closed trajectories. Our closely related approach allows also
for orbits that are not closed, but, of course, it can be applied also to
closed trajectories. This has been done in \cite{dm1}, where the
calculations of Kirzhnitz {\it et al.} have been repeated by using
the formalism discussed in this Section. In that study only one
``collective'' solution of the integral equation (\ref{pil}) was
found,moreover, thanks to the analytical insight given by the
semiclassical method, this solution was shown to be qualitatively different
from the plasma oscillations of a uniform electron gas. While plasma
oscillations belong to a strong-coupling regime (see e.g. \cite{fet},
p. 185), the ``collective'' solution found in \cite{dm1} pertains to a
regime of weak coupling. Thus it was concluded that
there is no evidence for plasmon-like excitations in this
semiclassical theory of atomic spectra. This conclusion is in
agreement with the result of analogous quantum calculations.

\subsection{Integrable spheroidal systems}

The equilibrium self-consistent mean field in heavy closed-shell
nuclei is well approximated by a spherical Saxon-Woods potential, with
parameters that can be determined phenomenologically. A somewhat
similar mean field describes also metal clusters \cite{eka}. Thus,
\begin{figure}
\centerline{\psfig{figure=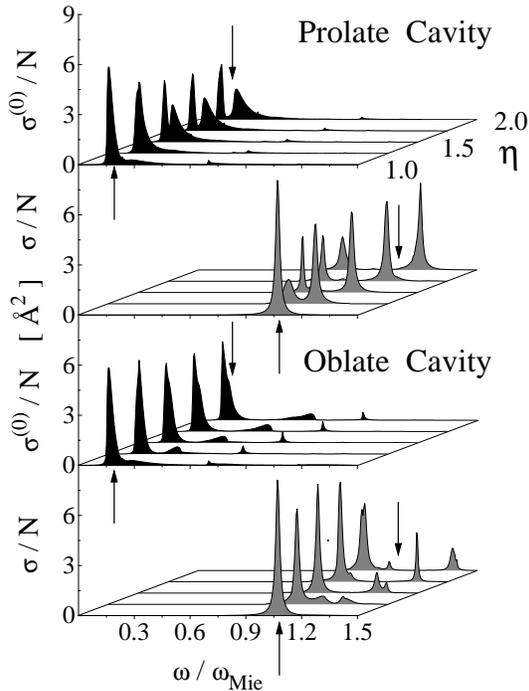,height=10.0cm}}
\caption{\small Photoabsorption cross section per valence electron (shaded
grey) for prolate and oblate sodium clusters. The black-shaded peaks show
results obtained in the single-particle approximation in which the
mean-field fluctuation (\ref{dev}) is neglected. The arrows indicate
the position of the plasmon peak in the spherical case.}
\end{figure}
for
open-shell systems it is reasonable to try a deformed Saxon-Woods like
potential. However the results of the first subsection cannot be
applied to this kind of potentials, since accurate calculations of the
classical phase space for axially deformed Saxon-Woods potentials have
shown that it contains regions of chaoticity \cite{arv}. But, if the
surface diffuseness is neglected and a spheroidal cavity is assumed to
approximate the mean field of large deformed nuclei \cite{str} or
clusters, then that formalism can be applied directly. This has been
done in Ref.  \cite{del} to study the surface plasmons in deformed
atomic clusters.
The main result of that paper is that, under certain conditions,
oblate deformed clusters may display a more complex shape of the
surface plasmon peaks than predicted by the classical Mie theory
(three peaks instead of two).
This prediction has not yet been checked experimentally.

Fig. 3 shows essentially the single-particle (shaded black) and
collective (shaded grey) dipole strength functions for prolate and
oblate spheroidal clusters. The figure is taken from Ref. \cite{del}.
The deformation parameter $\eta$ is the ratio between the larger and
smaller radius of the spheroid. The frequency is expressed in units
of the Mie frequency $\omega_{Mie}$. In the lower panel the
fragmentation of the high-frequency peak is clearly visible for
deformations $\eta=1.25$ and $\eta=1.5$.

\section{MOVING \\
BOUNDARY}

\subsection{Connection with liquid drop model}

In heavy closed-shell nuclei the equilibrium mean field can be further
approximated by a spherical square-well potential of radius $R\approx
1.2 A^{\frac{1}{3}}{\rm fm}$. In this simplified picture several of the
quantities needed to evaluate the linear response of the system
can be expressed analytically, hence it gives a useful reference model.

When the surface diffusion is neglected, the boundary condition
(\ref{bc2}) corresponds to requiring that the particles collide
elastically with the
static nuclear surface $r_{2}=R$ (mirror-reflection condition).

In Ref. \cite{ads} it has been shown that the results of \cite{bri}
can be generalized to a model in which the sharp surface is allowed
to vibrate, simply by changing the boundary condition (\ref{bc2}).

If the radius of a spherical nucleus is allowed to vibrate according
to the usual liquid-drop model expression \cite{boh}
\beq
\label{lds}
R(\theta,\varphi,t)=R+\sum_{LM}\delta R_{LM}(t)
Y_{LM}(\theta,\varphi)\,,
\eeq
then the mirror-reflection condition (\ref{bc2}) is changed to
\beq
\label{msc}
\delta \tilde f^{L+}_{MN}(R)-\delta \tilde f^{L-}_{MN}(R)=2F'(\epsilon)i\omega
p_{r}\delta R_{LM}(\omega)\,,
\eeq
where $p_{r}$ is the magnitude of the radial component of the particle
momentum.
We put a tilde over the moving-surface solutions to distinguish them
from the untilded fixed-surface solutions. The boundary condition
(\ref{msc}) still corresponds to a mirror reflection, but in
the reference frame of the moving surface.

It can be easily checked by direct substitution that the following integral
equation
\bea
\label{flmng}
&&\delta \tilde{f}^{L\pm}_{MN}(\epsilon,\lambda,r,\omega)=\\
&&F'(\epsilon)\frac{e^{\pm i[\omega\tau(r)-N\gamma(r)]}}{
\sin[\omega\tau(R)-N\gamma(R)]}\omega p_{r}\delta R_{LM}(\omega)
\nonumber\\
&&+{\big \{}\int_{r_1}^{r}dr' \tilde{B}^{L\pm}_{MN}(\epsilon,\lambda\, r')
e^{\mp [i\omega\tau(r')-N\gamma(r')]}\nonumber\\
&&+ \tilde{C}^{L}_{MN}(\epsilon,\lambda,\omega){\big \}}
e^{\pm i[\omega\tau(r)-N\gamma (r)]}\nonumber\,,
\eea
with
\bea
&&\tilde{B}^{L\pm}_{MN}(\epsilon,\lambda,r,\omega)=\\
&&F'(\epsilon){\Big (}\frac{\partial}{\partial r}\pm
\frac{iN}{v_{r}(\epsilon,\lambda,r)}\frac{\lambda}{mr^{2}}{\Big)}\nonumber\\
&&{\Big [}\beta(\omega) Q_{LM}(r)+\delta
\tilde{V}_{LM}(r,\omega){\Big]}\nonumber\,,
\eea
\bea
\label{delvtillm}
&&\delta\tilde V_{LM}(r,\omega)=\\
&&\frac{8\pi^{2}}{2L+1}\sum_{N=-L}^{L}
{\Big |}Y_{LN}(\frac{\pi}{2},\frac{\pi}{2}){\Big |}^{2}\nonumber\\
&&\int d\epsilon \int d\lambda \lambda \int \frac{d
r'}{v_{r}(r')}v_{L}(r,r')\nonumber\\
&&{\big [}\delta \tilde f^{L+}_{MN}(\epsilon,\lambda,r',\omega)
+ \delta \tilde f^{L-}_{MN}(\epsilon,\lambda,r',\omega){\big ]}\nonumber
\eea
and \footnote{There is a misprint in Eq. (3.16) of Ref.\cite{adm},
the present expression is the correct one.}
\bea
\label{clmn}
&&\tilde{C}^{L}_{MN}(\epsilon,\lambda,\omega)=\\
&&{\Big \{}e^{i2[\omega \tau(R)-N\gamma(R)]}\nonumber\\
&&\int_{r_1}^{R} dr
\tilde{B}^{L+}_{MN}(\epsilon,\lambda,r)
e^{-i[\omega \tau(r)-N\gamma(r)]} \nonumber\\
&&- \int_{r_1}^{R} dr \tilde{B}^{L-}_{MN}(\epsilon,\lambda,r)
e^{i[\omega \tau(r)-N\gamma(r)]}{\Big \}}\nonumber\\
&&\times{\Big \{}1-e^{i2[\omega
\tau(R)-N\gamma(R)]}{\Big \}}^{-1}\nonumber
\eea
is equivalent to the system
\beq
\label{systil}
\frac{\partial}{\partial r}\delta \tilde f^{L\pm}_{MN}\mp A_{N}\delta
\tilde f^{L\pm}_{MN}=\tilde B^{L\pm}_{MN}
\eeq
with the boundary conditions (\ref{bc1}) and (\ref{msc}),
and that it respects the self-consistency condition (\ref{dev}).

Equations (\ref{flmng}-\ref{clmn}) give an implicit solution of the
linearized Vlasov equation with moving-surface boundary conditions.
They are more complicated than the corresponding fixed-surface
equations because they contain also the unknown collective
coordinates $\delta R_{LM}(\omega)$. In order to proceed further, we
need an additional relation that specifies these quantities. As shown
in \cite {ads}, this relation can be found within the liquid-drop model
if one recalls that, in that model, a change in the
curvature radius of the surface results in a change of the pressure
at the surface  given by (cf. Eq. (6A-57) of \cite{boh})
\beq
\label{delp}
\delta {\cal P}(R,\theta,\varphi,\omega)=\sum_{LM}C_{L}\frac{\delta
R_{LM}(\omega)}{R^{4}} Y_{LM}(\theta,\varphi)\,.
\eeq

The restoring force parameter  $C_{L}$ can be easily evaluated: if the
Coulomb repulsion between protons is neglected,
\beq
\label{rfp}
C_{L}=\sigma R^{2} (L-1)(L+2)\,,
\eeq
where $\sigma\approx 1{\rm MeV\, fm}^{-2}$ is the phenomenological
surface-tension parameter, while taking into account also the Coulomb
interaction changes somewhat this expression \cite{boh}.

The connection with kinetic theory is then made simply by noticing
that the surface pressure (\ref{delp}) is an appropriate diagonal
element of the tensor (\ref{pre}) (generalized to Fermi liquids, see e.
g. \cite{lip}):
\bea
\label{prr}
&&\delta{\cal P}(R,\theta,\varphi,\omega)=\lim_{r\to R}\\
&&\int d{\bf p} p_{r}
v_{r}{\big (} \delta \tilde  f({\bf r},{\bf p},\omega)-F'(\epsilon)\delta
\tilde V({\bf r},\omega){\big )}\nonumber\,.
\eea
This integral must be evaluated for $r<R$ and then we must let $r\to
R$ from inside.

Multiplying both (\ref{delp}) and (\ref{prr}) by
$Y_{LM}^{*}(\theta,\varphi)$ and
integrating
over the directions of ${\bf r}$, gives
\bea
\label{delrlm}
&&\delta R_{LM}(\omega)=\\
&&\frac{8\pi^{2}}{2L+1}\frac{R^{2}}{C_{L}}\sum_{N=-L}^{L}
{\Big |}Y_{LN}(\frac{\pi}{2},\frac{\pi}{2}){\Big |}^{2}\nonumber\\
&&\int d\epsilon \int d\lambda \lambda p_{r}
{\Big [}\delta \tilde f^{L}_{MN}(\epsilon,\lambda,R,\omega)\nonumber\\
&&-2F'(\epsilon)\delta \tilde V_{LM}(R,\omega){\Big ]}\nonumber\,,
\eea
with the definition
\beq
\delta \tilde f^{L}_{MN}\equiv
\delta \tilde f^{L+}_{MN}+
\delta \tilde f^{L-}_{MN}\,.
\eeq
Equation (\ref{delrlm}) relates the fluctuations of surface and
distribution-function in the approach of \cite{ads};
it is an implicit relation for $\delta R$,
since both $\delta \tilde f$ and $\delta \tilde V$ do depend on $\delta R$.

\subsection{Approximate solutions for surface vibrations}

In order to obtain an explicit expression for the collective
coordinates $\delta R_{LM}(\omega)$ we make the following
approximation:

\bea
&&\delta \tilde f^{L}_{MN}(\epsilon,\lambda,r,\omega)=
\delta f^{L}_{MN}(\epsilon,\lambda,r,\omega)\\
&&+2F'(\epsilon)\frac{\cos[\omega\tau(r)-N\gamma(r)]}{
\sin[\omega\tau(R)-N\gamma(R)]}\omega p_{r}\delta R_{LM}(\omega)\nonumber\,,
\eea
with $\delta {f}^{L}_{MN}(\epsilon,\lambda,r,\omega)$ the
fixed-surface solution. This approximation is based on the assumption
that, apart from the term proportional to $\delta R_{LM}$, which alone
fulfills the boundary condition (\ref{msc}), in the bulk of the
system the moving-surface solution does not differ too much from the
fixed-surface one.

Then, Eq. (\ref{delrlm}) gives:
\bea
\label{last}
&&\delta R_{LM}(\omega)=\\
&&{\Big \{}\frac{8\pi^{2}}{2L+1}\sum_{N=-L}^{L}
{\Big |}Y_{LN}(\frac{\pi}{2},\frac{\pi}{2}){\Big |}^{2}\nonumber\\
&&\int d\epsilon\int d\lambda\lambda p_{r}{\Big [}\delta
f^{L}_{MN}(\epsilon,\lambda,R,\omega)\nonumber\\
&&-2F'(\epsilon)\delta V_{LM}(R,\omega){\Big ]}{\Big \}}
{\Big \{}\frac{C_{L}}{R^{2}}\nonumber\\
&&-\frac{(4\pi)^{2}}{2L+1}
\sum_{N=-L}^{L}
{\Big |}Y_{LN}(\frac{\pi}{2},\frac{\pi}{2}){\Big |}^{2}
\int d\epsilon F'(\epsilon)\nonumber\\
&&\int d\lambda\lambda p_{r}
{\Big [}\omega p_{r}\cot[\omega\tau(R)-N\gamma(R)]-\nonumber\\
&&\alpha_{L}(R,\omega){\Big ]}
{\Big \}}^{-1}\nonumber\,,
\eea
with
\bea
&&\alpha_{L}(r,\omega)=\\
&&\frac{(4\pi)^{2}}{2L+1}m\sum_{N=-L}^{L}
{\Big |}Y_{LN}(\frac{\pi}{2},\frac{\pi}{2}){\Big |}^{2}
\nonumber\\
&&\int d\epsilon F'(\epsilon)\int d\lambda \lambda
\int dr' v_{L}(r,r')\,\nonumber\\
&&\frac{\cos[\omega\tau(r')-N\gamma(r')]}{\sin[\omega\tau(R)-N\gamma(R)]}\omega
\nonumber\,.
\eea

Equation (\ref{last}) is an interesting, although approximate, expression
for the
collective coordinates given by the approach of Ref. \cite{ads}. The
vanishing of the denominator in this expression determines the poles
of $\delta R_{LM}(\omega)$, hence the eigenfrequencies of the
surface vibrations  $\delta R_{LM}(t)$.
Note that, according to Eq. (\ref{last}), for $L\neq 1$ these
eigenfrequencies
depend on the surface tension, which is physically sound.

The eigenfrequencies determined by the vanishing of the denominator
in Eq. (\ref{last}) can be directly compared to the experimental
excitation energies of the corresponding
modes. Up to now this has been done only after making an additional
approximation: if the fixed-surface solution is approximated by its
zero-order expression $\delta f^{0\,L}_{MN}$, appropriate for a gas
of non-interacting particles, a corresponding
zero-order approximation is obtained for $\delta R_{LM}(\omega)$
\bea
\label{last0}
&&\delta R^{0}_{LM}(\omega)=\\
&&{\Big \{}\frac{8\pi^{2}}{2L+1}\sum_{N=-L}^{L}
{\Big |}Y_{LN}(\frac{\pi}{2},\frac{\pi}{2}){\Big |}^{2}\nonumber\\
&&\int d\epsilon \int d\lambda\lambda p_{r}{\Big [}
\delta f^{0\,L}_{MN}(\epsilon,\lambda,R,\omega){\Big ]}{\Big \}}
{\Big \{}\frac{C_{L}}{R^{2}}\nonumber\\
&&-\frac{(4\pi)^{2}}{2L+1}\sum_{N=-L}^{L}
{\Big |}Y_{LN}(\frac{\pi}{2},\frac{\pi}{2}){\Big |}^{2}
\omega \int d\epsilon  F'(\epsilon) \nonumber\\
&&\int d\lambda\lambda p^{2}_{r} {\Big [}\cot[\omega\tau(R)-N\gamma(R)]{\Big ]}
{\Big \}}^{-1} \nonumber\,.
\eea
Note that this approximation contains more dynamics than the
corresponding fixed-surface zero-order approximation, since the
interaction between particles is taken into account to some extent
through the phenomenological surface-tension parameter $\sigma$.

\subsection{Surface and compression modes}

It is interesting to compare the dynamics described by Eq. (\ref{last0})
with that of the liquid-drop model.

The vanishing of the denominator in Eq. (\ref{last0}) can be written as
\beq
\label{eigf}
\frac{C_{L}}{R^{2}}-\frac{\chi_{L}(\omega)}{R^{2}}=0\,,
\eeq
with the function $\chi_{L}(\omega)$ defined as in Ref. \cite{adm}:
\bea
&&\chi_{L}(\omega)=\\
&&\omega R^{2}\frac{(4\pi)^{2}}{2L+1}
\sum_{N=-L}^{L}
{\Big |}Y_{LN}(\frac{\pi}{2},\frac{\pi}{2}){\Big |}^{2}
 \nonumber\\
&&\int d\epsilon  F'(\epsilon)
\int d\lambda\lambda p^{2}_{r} \cot[\omega\tau(R)-N\gamma(R)]\nonumber\,.
\eea

The expression analogous to (\ref{eigf})  in the liquid-drop model
is
\beq
\label{ldm}
C_{L}-\omega^{2} D_{L}=0\,,
\eeq
or, allowing also for compression modes (cf. Eq. (6A-58) of
\cite{boh}),
\beq
\label{ldmc}
C_L -\omega^2 D_L\,{L\over R}\,{{j_L ({\omega\over u_c }R)}
\over{\partial\over \partial r}j_L ({\omega\over u_c
}r)\mid_{r=R}}=0\,,
\eeq
with $u_c $ the velocity of sound. This condition determines the
eigenfrequencies of the various modes in the liquid-drop model
\cite{boh}. These eigenfrequencies depend also on the
mass parameters $D_{L}$ that can be evaluated within that model, but
one has to make some assumption about the kind of flow occurring in
the system \cite{boh}.

In order to make a comparison between the solutions of (\ref{eigf}) and
(\ref{ldmc}) it would be desirable to have a simple analytical
expression for the function $\chi_{L}(\omega)$.
The pole expansion of the cotangent
\beq
\cot z=\sum^{\infty}_{n=-\infty}\frac{1}{z-n\pi}
\eeq
allows us to express $\chi_{L}(\omega)$ in a form that is more similar to
the $\Pi^{0}_{L}$
propagator (\ref{spi0}):
\bea
&&\chi_{L}(\omega)=\omega R^{2}\frac{(16\pi)}{2L+1}\\
&&\sum_{n=-\infty}^{\infty}
\sum_{N=-L}^{L}{\Big |}Y_{LN}(\frac{\pi}{2},\frac{\pi}{2}){\Big |}^{2}
\int d\epsilon  F'(\epsilon)\nonumber\\
&&\int d\lambda\lambda
p^{2}_{r}\frac{\omega_{0}}{(\omega+i\varepsilon)-
(n\omega_{0}+N\omega_{\varphi})}\nonumber\,,
\eea
but this only shows that $\chi_{L}(\omega)$ is a rather involved
complex function. The very fact that $\chi_{L}$ is complex is already
an interesting result since we can expect complex eigenfrequencies as
solutions of Eq. (\ref{eigf}), in close analogy with the Landau
damping phenomenon in homogeneous system and in contrast to the
liquid-drop model condition (\ref{ldmc}) that involves only real
quantities.

In the case
$L=0$ the function $\chi_{L}(\omega)$  can be evaluated analytically
\cite{ads},
while for $L=1,2,3$ relatively simple expressions have been obtained
for the coefficients of the low-frequency expansion
\beq
\chi_{L}(\omega)\approx A_{L} +i\omega\gamma_{L}+D_{L} \omega^{2}\,.
\eeq
For isoscalar monopole vibrations, the solution of Eq.(\ref{eigf})
compares reasonably well with the experimental value of the monopole
giant resonance energy in heavy nuclei.

For isoscalar quadrupole oscillations, in \cite{adm} it was concluded that the
approximation (\ref{last0}) is not adequate to describe the low-energy
isoscalar
quadrupole excitations. It is not clear if this failure is
due to the neglect of the bulk mean-field fluctuations in (\ref{last0}) or
to some other reason. Further work is needed on this problem.

About the low-energy isoscalar octupole oscillations, in \cite{adm} it was
concluded that they are overdamped in the approximation given by Eq.
(\ref{last0}), and this might offer a qualitative explanation for the
background observed in inelastic proton scattering \cite{ara}.

Another application of Eq. (\ref{last0}) has been made in Ref.
\cite{adm2}. There this equation has been used to study the isoscalar
dipole compression mode. In order to get reasonable results in this
case it is vital to treat the centre of mass motion correctly, and
this can be done in a fully self-consistent approach. Because
Eq.(\ref{last0}) is not really self-consistent, in \cite{adm2} it was
found necessary to add an extra term to the dipole response function
based on Eq. (\ref{last0}). For other multipolarities this extra term
can also be interpreted
as the contribution to the response due to the change of shape
of the system in the moving-surface approach, for $L=1$ the
system does not change its shape, but can translate as a whole. Since
there is no force opposing this translation, this centre of mass
motion results in a pole at $\omega=0$ in the dipole response function.
The interesting physics is contained in the intrinsic response, that
is shown in Fig. 4.
\begin{figure}
\centerline{\psfig{figure=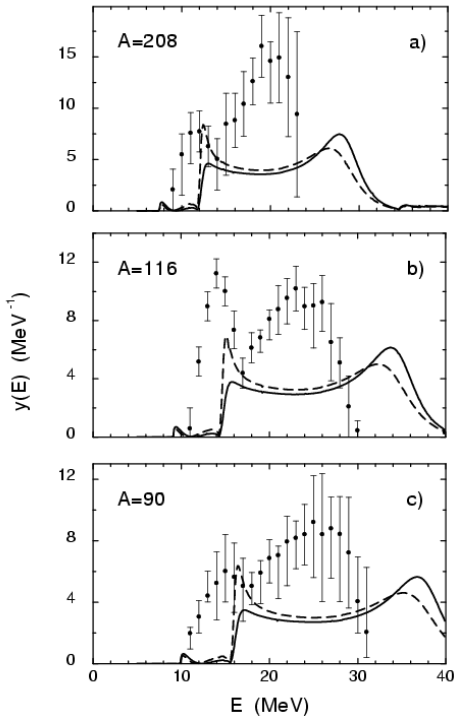,height=10.0cm}}
\caption{Intrinsic dipole energy-weighted strength
function compared to experimental data for $^{208}{\rm
Pb}$ (a), $^{116}{\rm Sn}$ (b) and $^{90}{\rm Zr}$ (c). The solid curve
is for a confined gas of $A$ non-interacting nucleons with  incompressibility
$K\approx 200 {\rm MeV}$, the dashed curve
is for interacting nucleons with $K=160 {\rm MeV}$. The data are
from Ref.\cite{you}}
\end{figure}

The results of the moving-surface theory are in qualitative agreement with
experimental data on the isoscalar dipole mode, as can be seen from Fig.
4. Moreover, the insight
given by this semiclassical approach gives us the possibility
to establish a direct link between the peak profile of the isoscalar dipole
resonance and the incommpressibility of nuclear matter \cite {adm2}.
This link is evidenced by the dashed curve in Fig. 4.

The approach described here can be easily generalized to a
two-component fluid. In \cite{abro} such a generalization has been
used to study the effect of neutron excess on the isoscalar and
isovector monopole response functions. For such a two-component
system, Eq.(\ref{prr}) is replaced by the two following equations
\bea
&&\delta{\cal P}(R,\theta,\varphi,\omega)=\\
&&\Pi_{rr}^n (R,\theta,\varphi,\omega))
+\Pi_{rr}^p(R,\theta,\varphi,\omega)\nonumber\,
\eea
and
\bea
&&\delta{\cal P}_{Q}(R,\theta,\varphi,\omega)=\\
&&\Pi_{rr}^n (R,\theta,\varphi,\omega))
-\Pi_{rr}^p(R,\theta,\varphi,\omega)\nonumber\,.
\eea
Here $\Pi_{rr}^{n(p)}(R,\theta,\varphi,\omega)$ is the normal component of the
momentum-flux tensor  [r.h.s. of Eq. (\ref{prr})] associated with neutrons
(protons) only. Apart from the usual surface pressure $\delta{\cal P}$ that
is present when
the neutron and proton surfaces move in phase, in this case there is
also an additional pressure $\delta{\cal P}_{Q}$ that is caused by
the forces opposing the pulling apart of the neutron and proton
surfaces when they move out of phase. This extra pressure can also be
related to an appropriate phenomenological parameter.
With the surface energy of \cite{swiat}, the pressure $\delta{\cal P}_{Q}$
can be written as
\bea
\label{delpiv}
&&\delta {\cal P}_{Q}(R,\theta,\varphi,\omega)=\\
&&{Q \over {2 \pi r_0^4}}
\left[\delta R_n(\theta,\varphi,\omega)-
\delta R_p(\theta,\varphi,\omega)\right]\nonumber\,,
\eea
where $Q$ is the neutron skin stiffness coefficient of \cite{myers}
that is analogous to the surface tension parameter $\sigma$
of Eq. (\ref{rfp}) and
$r_{0}=1.14\,{\rm fm}$.

As shown in  \cite{abro}, the neutron excess does not affect much the
isoscalar giant monopole resonance, while its effect on the isovector
resonance is more pronounced.

\subsection{Fixed vs. moving boundary}

The possibility of solving the Vlasov equation with different
boundary conditions allows us to decide which condition is more
appropriate by comparison with experiment. In Fig. 5 we give an
example of a nuclear response function evaluated according to the
fixed- and moving-surface boundary conditions.
\vskip 0.5cm
\begin{figure}
\centerline{\psfig{figure=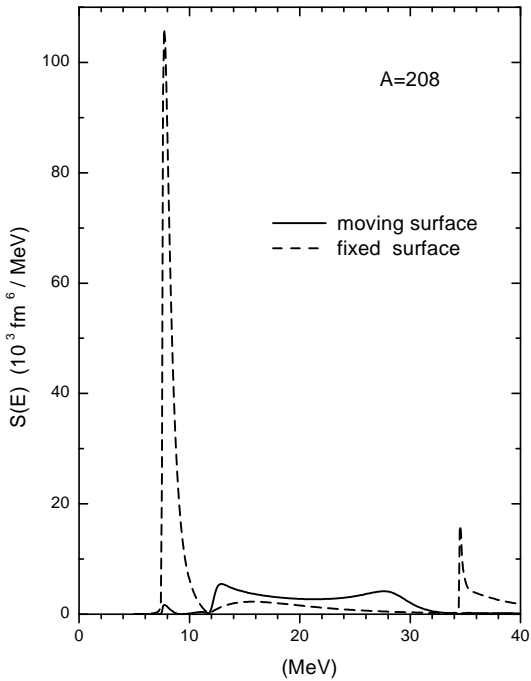,height=9.0cm}}
\caption{Isoscalar dipole strength functions evaluated in zero-order
approximation with moving-surface (full curve) and fixed-surface
(dashed curve) boundary conditions. The full curve corresponds to the
full curve in Fig.4}
\end{figure}
In this case there is
no doubt that comparison with data (shown in Fig. 4) favours the
moving-boundary solution, at least at the level of zero-order
approximations. The present example is a rather extreme case, for
other response functions the difference between the two solutions is
not expected to be so large. Work on detailed comparison between the
two solutions for different multipolarities is still in progress.

\section{QUANTUM EFFECTS}

The Vlasov theory is completely classical, however it is easy to
establish a connection with quantum mechanics. It is well known that
the time-dependent Hartree-Fock equation for the Wigner transform of
the quantal one-body density matrix reads (see e.g. Ref. \cite{rin},
p. 553)
\bea
\label{tdhf}
&&\frac{\partial f({\bf r},{\bf p},t)}{\partial t} +
\frac{2}{\hbar}f({\bf r},{\bf p},t)\\
&&\times \sin[\frac{1}{2}\hbar({\overleftarrow{\bf
\nabla}}\cdot{\overrightarrow {\bf \nabla}}_{p} -
{\overleftarrow {\bf \nabla}}_{p}\cdot{\overrightarrow {\bf
\nabla}})]\: [\frac{p^{2}}{2m}\nonumber\\
&&+V({\bf r},{\bf p},t)]=0 \nonumber
\eea
and reduces to
\bea
\label{cl}
&&\frac{\partial f}{\partial t} + \frac{\bf p}{m}\frac{\partial
f}{\partial {\bf r}} \\
&&-(\frac{\partial V}{\partial {\bf r}})\cdot\frac{\partial
f}{\partial {\bf p}}+ (\frac{\partial V}{\partial {\bf p}})\cdot\frac{\partial
f}{\partial {\bf r}}=0 \nonumber
\eea
if $\hbar\to 0$. The Hartree-Fock potential $V({\bf r}, {\bf p},t)$
is momentum-dependent because the Fock term is non-local in coordinate
space. If the momentum-dependence of $V$ is neglected (Hartree
approximation), then Eq. (\ref{cl}) coincides with Eq. (\ref{kin}).

There are several possibilities for including quantum effects into
the Vlasov theory. The most trivial one is to choose an appropriate
equilibrium distribution $F(h_{0}({\bf r}, {\bf p}))$. The standard
choice for zero-temperature fermions is
\beq
\label{f0}
F(h_{0}({\bf r}, {\bf p}))=\frac{g}{(2\pi\hbar)^{3}}\vartheta
(\epsilon_{F}- h_{0}({\bf r}, {\bf p}))\,,
\eeq
where $\epsilon_{F}$ is the Fermi enegy and $g$ is the degeneracy
factor ($g=2\,(4)$ for electrons (nucleons)).

The choice (\ref{f0}) respects the Pauli principle in the sense that
there cannot be more than one fermion of given spin (and isospin) in
a phase-space cell of volume $h^{3}$. As stressed by Bertsch
\cite{ber}, the Liouville theorem then ensures that the Pauli principle
will be respected during the time evolution of the system described
by the Vlasov equation. Thus the Liouville theorem, stating that the
points representing the system in the 6-dimensional phase space evolve
as an incompressible fluid, justifies the use of the Vlasov equation
for quantum systems. Using the equilibrium distribution (\ref{f0}) makes
the completey classical Vlasov approach a semiclassical one.
Clearly in this way we do not expect to
reproduce details, but only gross properties of quantum many-body
systems.

In principle one could introduce further quantum effects also by taking into
account terms of order $\hbar^{3}$ and higher that have been
neglected in Eq. (\ref{cl}). The problem with such a direct approach
is that the semiclassical calculations become quickly more complicated than
the corresponding quantum calculations, so this method would not be very
convenient. A more pragmatic attitude has been taken in
Refs. \cite{dm2} and \cite{abm}, where the semiclassical propagators
given by the Vlasov theory have been compared directly with the
analogous quantum propagators.

In \cite{dm2} the relation between the
Fourier coefficients (\ref{fou}) and the corresponding quantum matrix elements
has been discussed. If single-particle matrix elements are
evaluated in the WKB approximation, they tend to Fourier coefficients
analogous to (\ref{fou}) in the limit of large quantum numbers, but
this is not necessarily the case for the exact quantum matrix elements.

In \cite{abm} the closed-orbit theory, that had
been used by various authors to evaluate quantum corrections to the
Thomas-Fermi level density, has been applied to evalute quantum
corrections to the Vlasov propagator (\ref{qpi0}). The resulting
strength function compares better with the analogous quantum strength
function. A fascinating aspect of this study is the possibility of
relating features of the excitation spectrum of quantum systems to
simple properties of the classical orbits.

\section{CONCLUSIONS}

 In the kinetic-theory approach to linear response
discussed here the finite size of the system has been explicitly taken
into account. Two different boundary conditions for the density
oscillations have been considered: fixed and moving surface.

The fixed-surface solution is closely related to the quantum random
phase approximation (RPA) since the integral equation (\ref{rpa}) for
the semiclassical collective propagator is essentially the same as
the equation for the quantum RPA propagator when exchange terms are
neglected (Hartree approximation). The main difference between the
semiclassical and quantum approaches  is in the ``free'' propagator
$\Pi^{0}$. While the quantum expression for this propagator involves
wave functions  and
single-particle energies, the semiclassical propagator involves only
properties of the classical orbits and is considerably simpler to
evaluate.

The moving-surface solution has been related to the liquid-drop model
of nuclei. Compared to that model, the present approach is more
microscopic and it allows also for the possibility of Landau damping
of collective vibrations.

When compared to experiment, in some cases the semiclassical solutions have
the
same problems of the RPA and of the liquid-drop model.
For example, the fixed-surface approach is, like RPA, able to describe
reasonably well the position of the surface plasmon in atomic
clusters, but, again like RPA, fails to reproduce the observed width
of this resonance. Possible moving-surface contributions to this
width have not yet been investigated in detail within the present
approach. The moving-surface approach shares with the liquid-drop model
the difficulties in reproducing the experimental value of low-energy
quadrupole oscillations in heavy nuclei. However the insight given by
this approach could be useful for improving both the quantum RPA
theory and the liquid-drop model.
In conclusion, the semiclassical theory discussed here offers an
interesting alternative to the quantum RPA approach because it is
simpler to implement numerically, but it is still sufficiently
sophisticated to take into account some realistic features (finite
size, equilibrium shape) of the many-body system.
\section*{ACKNOWLEDGMENTS}
We are grateful to Prof. D.M. Brink for his careful reading of the
manuscript.

\end{document}